\documentclass[prb,preprint,floatfix]{revtex4-1} 

\usepackage{amsmath}  
\usepackage{amstext}
\usepackage{amsfonts} 
\usepackage{graphicx} 

\begin{document}


\title{Predicting the next local supernova}

\author{John Middleditch}
\email{j.middleditch@gmail.com}
\affiliation{University of California, retired}

\date{\today}

\begin{abstract}

Core collapse within blue supergiant stars, as occurred within Sk 
-69$^{\circ}$202/Supernova 1987A, is generally attributed to a 
merger of two electron-degenerate cores within a common envelope, 
with a merged mass in excess of 1.4 solar.  Supernova 1987A also 
had \textit{two} associated bright sources, one with about 8\% of 
the H$\alpha$ flux, and 74 milli-arc seconds distant by day 50, 
and another, four times fainter and 160 milli-arc seconds away 
in the opposite direction on day 38, when the first bright source
was only 60 mas distant. Using recent advances in 
our understanding of pulsars, we can show that the second source 
was the result of the core-merger process, which can drive a 
relativistic jet of particles \textit{prior} to the completion 
of the merger process, \textit{whether this proceeds to core 
collapse, or not}.  As with those resulting from core-collapse, 
such beams and jets are likely to produce an obvious spectral 
signature (e.g., even in \textit{un}-red/blueshifted H$\alpha$), 
which can be detected in nearby galaxies.  There is very likely 
a time interval of a few months, during which such supergiant 
stars, a high fraction of which will eventually undergo core 
collapse, can be identified.  These can be carefully followed 
observationally to maximize the chance of observing core 
collapses as they happen. Such studies may eventually help in 
using such objects as standard candles.
\end{abstract}

\maketitle 

\section{Introduction} 
\label{sec:intro}

As recently as 1987, orthodox supernova theory did not anticipate
that a star could undergo core collapse while it was a blue supergiant 
-- only during the supposedly later red supergiant phase was this 
expected.  On 23 February, 1987, 
Supernova 1987A in the Large Magellanic Cloud, the offspring of the 
blue supergiant star, Sk -69$^{\circ}$202, changed that expectation.  
The simplest explanation for this is that Sk -69$^{\circ}$202 was, in 
fact, a star with two electron degenerate stellar cores within a
common envelope (in isolation these would be white dwarf stars) which 
were about to merge.\cite{PJ89,P91}  The observation of a 
2.14 ms signal from SN 1987A in the optical/near-infrared, over a 
timespan of four years, is consistent with this process, as is the 0.62 
foe energy drop from an initialspin period of 2.00 
ms.\cite{M00a,M00c,M18}

The merger was the result of a binary stellar system close enough 
so that the friction of the motion of the binary components within 
the expanding stellar envelopes was sufficient to cause the orbits 
of the cores to decay, moving closer to their 
companion with time.  There is plenty of other evidence for 
binarity in the progenitor, including the rings,\cite{C95,MP07}
and the mixing observed in the ejecta,\cite{C87,HD87}
but we note in [\onlinecite{M18}], that the anisotropy of 
the expanding remnant\cite{Pap89,W02} is more a result of 
the SN disruption process.

\section{The observations}

A bright source (BS1) only 45 milli-arc s (mas) from, and 
amounting to 8\% of, the flux of SN 1987A, was 
observed\cite{Nis87} on day 30 in a 10 nm-wide filter 
overlapping H${\alpha}$,
at a bearing of 194$^{\circ}$. By day 38 it was 60 mas away, 
and on day 50, 74 mas distant (still at the same bearing, 
and days 29.8, 37.8, and 49.8  are used for purposes 
of calculation).\cite{MMM} 
The source was not detected when it was observed again on 
day 98 by the first group.  However an improved analysis on 
data from day 38, done by them a decade later,\cite{NP99} 
detected another nearby bright source (BS2), four times 
fainter than the first (at magnitude 8.25, still the 
brightest source in the LMC, other than SN 1987A proper 
and BS1), and 160 mas away \textit{in the opposite 
direction -- the line joining the two contained SN 1987A
proper}.

The BS1 data, along with the early light curve of SN 1987A 
from the Cerro Tololo Inter-American Observatory (CTIO) 
24-inch telescope\cite{HS90} 
and the International Ultraviolet Explorer,\cite{W87} 
were used to solve for the offset and depth of circumpolar 
ejecta near Sk -69$^{\circ}$202, and the orientation and 
kinetics of a beam of radiation and jet of particles 
emerging from the star's (south) polar direction.\cite{M12}

The solution revealed a light beam and particle jet with 
a collimation factor $<10^{-4}$, an orientation of 
75$^{\circ}$.2 (slightly pointed toward Earth, but mostly 
south), and 10.5/25.5 light-days ($\ell$t-d) to the 
beginning/end of the circumstellar ejecta above the south 
pole of Sk -69$^{\circ}$202.  The fastest particles in the 
jet travel at almost 0.958 c, which is 
consistent with the kinetic energy of the peak in the 
proton cosmic ray spectrum near 2.--2.5 GeV. 

The only way that both BS1 and BS2 could have been the 
result of jets driven by post-core-collapse Sk 
-69$^{\circ}$202 (SN 1987A), was for BS2 to be 
approaching, in order that it not be overwhelmingly 
delayed, as was first noted by [\onlinecite{NP99}],
which means the two sources and 87A would not lie on a 
single line. However, Occam's razor would suggest that 
if BS1, 87A, and BS2 appear to be colinear in the 
plane of the sky, then it is most probable that they 
are, in fact, colinear in three dimensions, so that BS2 
is slightly receding.  Moreover, the solution to the 
geometry as done in [\onlinecite{M12}] \textit{assuming} 
BS2 came from the core collapse, as is the case for BS1, 
has BS1 emerging at 62.4$^{\circ}$, still 12.8$^{\circ}$ 
off the 75.2$^{\circ}$ solution from [\onlinecite{M12}], 
with a peak of the proton cosmic ray spectrum well beyond 
2.5 GeV (Fig.~8 of [17]), in conflict with observation.  

Finally, the right triangle at day, $d=37.8$, made 
from the approaching 
solution BS2, measured at $h$ (45.93 $\ell$t-d,
or 160 mas) from 87A at (0,0), and ($\rho - d$, 0),
has a hypotenuse of $\beta \rho$.
With $\rho$ as the current radius of the first SN 
flash, and $\beta$c the mean velocity for the matter 
swept up to form BS2, a real solution for $\rho$ 
occurs only if $\beta> h/\sqrt(h^2+d^2) = 0.772$, 
which is highly unlikely.  

Even a beam of light from the SN flash, which hit 
matter at BS2 in time to illuminate it at day 37.8, 
still requires ejection and subsequent ``sweepup'' 
of material, \textit{prior} to core collapse, to 
form the target the beam eventually hits. It is 
also unlikely that scattered light alone can make 
BS2 as bright as observed. For all of these 
reasons, BS2 can not possibly have originated 
from the core-collapse event.  Figure 1 shows a 
geometry of the system (from [\onlinecite{M12}]),
assuming BS2, 87A, and BS1 are, in fact, colinear 
in three dimensions (but was calculated for day 
30 instead of day 38).

\begin{figure}[ht!]
\centering
\includegraphics[height=14cm,width=20.3814cm, 
trim={1.5cm 2cm 0cm 1.1cm}]{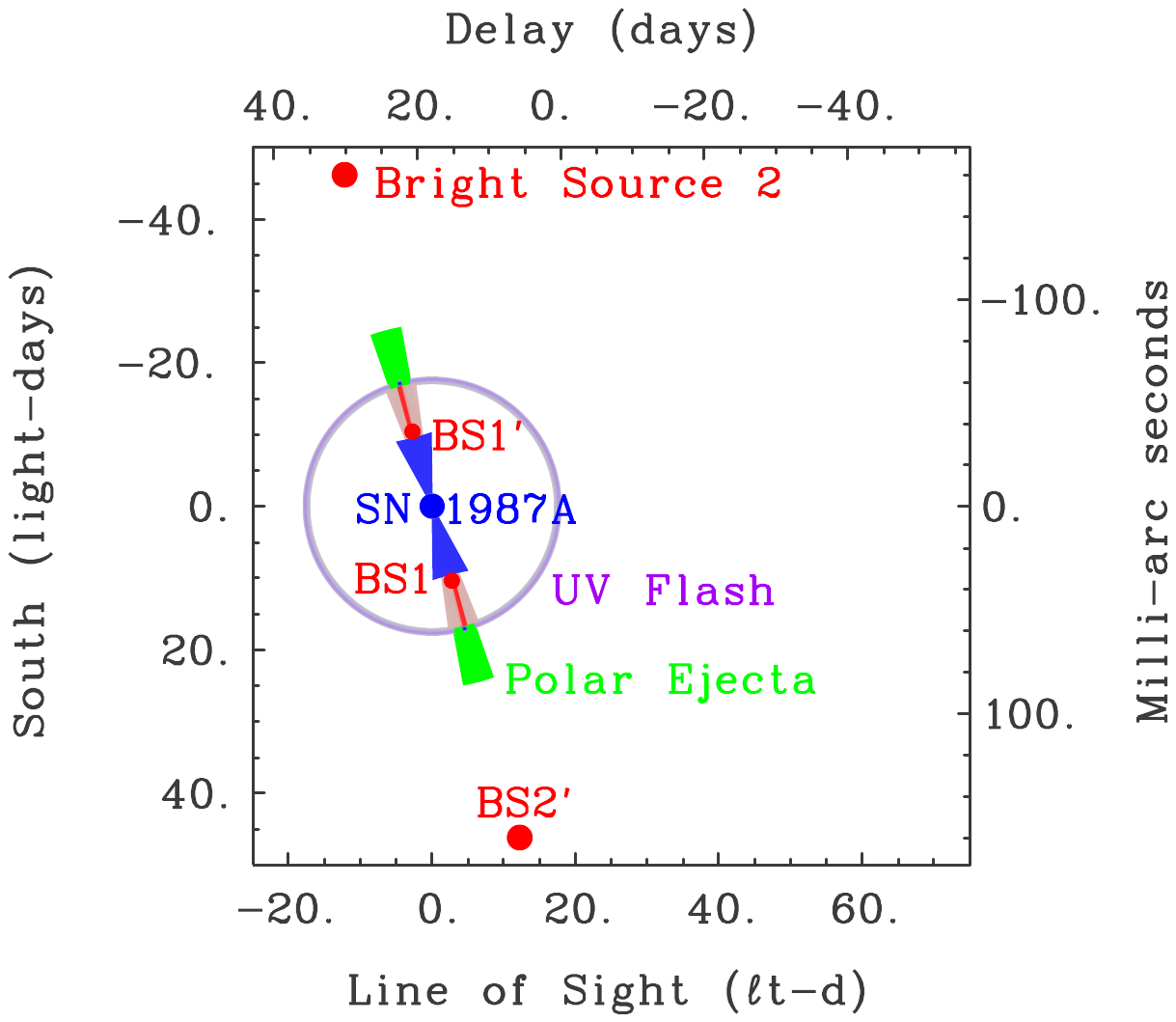}
\caption{The geometry of SN 1987A on day 18 and associated 
sources (at other corresponding times -- [\onlinecite{M12}]).
Bright Source 2 on day 38 (this figure is for day 
30 -- top horizontal scale),
was observed to be 160 mas north of SN 1987A (right
hand vertical scale).  The forward counterpart of Bright 
Source 2, BS$2\sp{\prime}$, would have been observed to 
be 160 mas south of SN 1987A on day 14 (again top scale
-- this date has been corrected for BS2 at day 38).  
BS$1\sp{\prime}$, the back counterpart of BS1, at day 21 
has been arbitrarily set at 36 mas north of SN 1987A 
(right scale).  BS1 at day 15 and BS$1\sp{\prime}$ 
are just starting to form from the beam and jet (red) 
impacting the beginning of the polar ejecta (green, 
starting at $\pm10.5~\ell$t-d, but graphically 
obscured by ejecta until after $\pm 17~\ell$t-d in 
projection).\label{fig:BS12}}
\end{figure}

Opposite sources for BS1 and BS2 have been added to 
Fig.~1. These have never been observed, but for good 
reasons. Source BS$1\sp{\prime}$ suffers from greater
opacity from the circumstellar material than does BS1.  
The breakout of the beam of light,\cite{Men87}
i.e., the ``Bochum Event'' at day 19.2 (Fig.~5 (b) `E' 
of [\onlinecite{M12}])
attests to the significance of this opacity.

The counterpart to Bright Source 2, BS$2\sp{\prime}$, would 
have been exactly opposite BS2 on day 14 (top scale of Fig.~1,
plus 8 to compensate for the time difference to day 38).  
The speckle observation sufficiently sensitive to detect it, 
however, was made on day 38 so BS$2\sp{\prime}$ would have 
suffered 16 more days of fading and motion away from 87A 
(3.2 more $\ell$t-d at 0.2 c), by then, in all likelihood, 
invisible to any analysis.  Again from Fig.~1, even corrected
to day 38, it is obvious that Bright Source 2 was a result of 
a polar beam of light and jet of particles ejected from Sk 
-69$^{\circ}$202 -- the progenitor \textit{star} -- just 
as BS1 was a similar result of SN 1987A.

\section{Interpretation}

Recent theoretical advances in our understanding of the 
pulsar mechanism\cite{Ar94,H98,AR03,AR04,AR07,AR08}
have explained how polarization currents, induced well 
beyond the pulsar's speed-of-light cylinder (and so 
updated in a \textit{pattern} much faster than light), 
and within the progenitor's plasma, will drive a 
highly collimated ($<10^{-4}$) beam of light and jet of 
particles in the polar directions.  Each annulus of 
polarization currents, coaxial with the star's 
rotation axis, produces two focused beams, whose paths 
are given by
\begin{equation}
\label{Z(XR);Y(R)}
Z =\pm  \sqrt{(R^2 - 1)(X^2 - 1 + \frac{1}{R^2} )};~Y=1/R;~R \ge 1
\end{equation}
for a current source at (0,$R$,0), on an annulus in 
the X-Y plane centered at (0,0,0), where $R$ is the 
ratio of the polarization pattern update speed (in the 
+X direction) to c, and $X$, $Y$, 
and $Z$ are measured in light-radians of the pulsar.  
Thus the beams propagate in an X-Z plane with a 
Y-intercept of $1/R$, and their power 
\textit{diminishes only as 1/distance}.  The larger 
$R$ is, the more polar the beam. Typically, $R$ 
exceeds 100,000 toward the equators of blue 
supergiants disrupted by a 500 Hz pulsar, and 
likewise for red supergiants disrupted by a 50 
Hz pulsar (see immediately below).  The asymptotic 
pattern of the focused radiation, from an annulus of 
current sources in the X-Y plane, is two circles on 
the sky of equal and opposite spin latitudes 
given by $\pm \arccos(\mathrm{c}/v)$, where $v$ is 
the update speed of the polarization current 
pattern.\cite{M18}

Just as a rotating
neutron star can excite polarization currents in an annulus
with a pattern which is updated faster than c, so can two 
co-orbiting C-O white dwarf cores.  Their magnetic field lines 
will thread and behave, more or less (with spin as a wild card),
as if the two cores were a single rotating, but not necessarily
aligned, magnetic dipole.  If we let 
$r_{\mathrm{eq}}$ represent the equatorial radius of Sk 
-69$^{\circ}202$, $a$ the separation of two 0.7 
M$_{\bigodot}$ electron degenerate cores, and $P$ their orbital
period, ignoring the other material in the stellar interior, we 
have, for $R_{\mathrm{eq}}$, the ratio of the equatorial 
excitation speed, $v_{\mathrm{eq}}$, to c:
\begin{equation}
\label{Kepler}
R_{\mathrm{eq}}~\equiv~v_{\mathrm{eq}}/\mathrm{c}~=~2 \pi 
{{r_{\mathrm{eq}}/(P\mathrm{c})} 
=~0.4547~\frac{{(r_{\mathrm{eq}}/
(10^7~\mathrm{km})) }}
    {(a/(10^5~\mathrm{km}))^{3/2}}}~.
\end{equation}

The equations for $P$ and $a$ are useful to have on hand:
\begin{equation}
\label{Pa}
a=59,131~\mathrm{km} \times(r_{\mathrm{eq}}/(R_{\mathrm{eq}}~
10^7~\mathrm{km}))^{2/3}~;~~P~=~460.9~\mathrm{s}
\times(a/(10^5~\mathrm{km}))^{3/2} ~.
\end{equation}
The equatorial excitation velocity reaches c ($R_{\mathrm{eq}} = 
1$) for $r_{\mathrm{eq}}/(10^7~\mathrm{km}) = 2$, 
$a = 93,864~\mathrm{km}$, and when the orbital period is 419 s 
(also the circumference of the star in $\ell$t-s).

Since much of the non-degenerate stellar core will rotate with 
the two cores, the separation between the two may not be in 
accordance with Eq.~\ref{Kepler}.  Also in order for the
beam and jet to be well collimated, we need $R >> 1$.  Although
the $2 \times 10^7$ km radius of the blue supergiant\cite{BW20}
limits R to be less than 210, assuming a minimum orbital period
of 2 s ($a$=2,660 km), this is still sufficient to drive
a highly collimated beam and jet, after first driving less
collimated features (there is no precession for this
process, unlike post-core collapse).

It is unclear how much of the shroud material, experienced
by the post-core-collapse beam and jet, is due to
the core-merger process.
This complicates the estimate of the time delay between
the appearance of BS2 and core collapse.  If BS2 encountered
as much circumstellar material as BS1 did, then, judging 
from Fig.~1, getting to 74 mas (all projected from 
$75.2^{\circ}$) within the polar ejecta took 50 days.  There 
remains another 10 mas or 2.9 $\ell$t-d of polar ejecta to 
get through, and then another 21 $\ell$t-d at something 
near 0.2 c, which amounts about 171 days total (deprojecting
all but the 50 days).  

The continued slope\cite{M18} of the early
light curve after the Bocum Event suggests that there 
may still have been material to be swept up beyond 25 
$\ell$t-d, so this ``coasting'' velocity may have fallen
below 0.2 c.  On the other hand, in the unlikely
possibility that there is little polar ejecta to ``bunch up,'' 
until a target of material already at 160 mas (projected),
then the extra time between merger beam and core-collapse
beam would only be 25 $\ell$t-d, which deprojects to a 
25.86-day interval.  The physical reality is likely to be 
some time interval between the extremes, and five
months might be a good guess. Of course this only applies 
to the geometry of SN 1987A -- the delays to core-collapse
of other systems with less oblique geometries could be
much smaller.

\section{discussion}

Normally, one would not expect that a large star, with two tiny
cores churning deep within its center at what one would think 
was a lazy pace, could drive relativistic jets of particles 
from its poles, but that's how the mathematics work, and in 
consequence, how the Universe must work (the focusing 
calculated in [\onlinecite{M18}] is so extreme that the 
defocusing effects of the stellar interior are not
likely to change this).  In fact, 
\textit{all}\cite{M18} large stars will eventually produce 
polar jets as they develop rotation and magnetic fields 
within their core(s), whose moments of inertia drop with 
time due to orbital decay for doubles, and burning to 
heavier elements for massive singles.


Although the velocities of 
C or O ejected from stars without H or He, the progenitors of 
SNe Ia, will be less 
than the 0.957 c of H, they need to be at least 0.75 c
if they drove the H in SN 1987A by collision, and 0.8747 c
if they were all driven electromagnetically, with equal 
energy per proton, assuming nuclei with all of their 
companion electrons ionized away.  However, the H driven
by the pulsar companion to SS 433 is line-locked at just
under 0.28 c, probably due to the absence of heavy 
elements.\cite{M12}

It is unknown whether there were any recognizable H$\alpha$
lines with up to 40\% red/blueshifts associated with SN 1987A
(the 8\% in H$\alpha$ is limited to close to $\sim$6561$~
\mathrm{\AA}$, but there is still H$\alpha$ flux well outside 
of this band because there are also parts of BS1 with its bulk, 
particle beam, and target impact velocities, among others).  
However the main focus back then was on a 
continuum of smaller shifts.  Sk -69$^{\circ}$202, 50 kpc 
distant in the LMC, with a B magnitude of 12.28, would be 
18.8 magnitudes at 1 Mpc, still enough light to get a 
spectrum with a big telescope.

The entire local group lies within 1 Mpc, including both M31 
and M33.  NGC 55, 300, and 3031 (M81) lie at 1.2, 1.3, and 
1.4 Mpc respectively, and the blue supergiants in all three 
are still accessible.  Perhaps the most interesting 
possibility is the ability to predict core collapse to a 
black hole, and the more massive, and certainly brighter 
supergiants likely to do this would be accessible out 
to 5 Mpc, which would bring NGC 253, 1313, 2403, 3034 
(M82), 4449, 4736 (M94), 4826 (M64), 4945, 5128 (Cen A), 
5236 (M83), 5457 (M101), 6946, and 7793 into range.

Red giants and supergiants, however, will likely have polar
jets well in advance of their impending core collapse.  In 
such stars of 20 to 25 solar masses, oxygen burning will 
last a handful of months, while silicon burning will last a 
handful of days.  In order to predict their core-collapse 
events in good time, we will need a way to recognize the 
transition to oxygen burning.

Aside from 
emission characteristic of a strongly
magnetized young pulsar remnant in VT 1137-0337,\cite{DH23} 
no such central source has been detected 
in any other SN. 
This may mean that 1993J, although a SN of a red supergiant, 
was still due to a merger (but one with an extreme mass 
ratio) so that any remnant would not be strongly magnetized. 
The merger may have left a black hole, but it is not clear 
whether this would have led to a recognizable SN, as was
observed. However, even such a merger within a red supergiant 
should still produce high velocity jets in the 
pre-core-collapse phase. 

\section{Conclusion}

The second bright source near SN 1987A, BS2, found by 
[\onlinecite{NP99}]
can only be successfully interpreted as a mildly 
relativistic jet of particles (and likely a beam of 
light) driven by the core merger process 
\textit{months} prior to core collapse.  
These effects may
allow the spectral identification of Galactic and 
extragalactic supergiants which will undergo core 
collapse within a few months, leading to the exciting 
possibility of observing core collapse as it happens.  
In the long term we will need to observe many more 
such events in the hopes that these may eventually 
be used as standard candles, which, at present, is 
impossible.\cite{M18} There are also those events in 
the brightest supergiants leading to black holes, 
about which almost nothing is known. Given the new 
understanding of disruption from core-collapse 
developed in [\onlinecite{M18}], it is not at all 
clear what such events would look like, as there 
would be no pulsar to disrupt the star.

\section{Addendum}

Observations of the progenitor of SN 2020tlf reported in 
[\onlinecite{jg22}] showed a roughly constant jump in 
luminosity 
for 130 days prior to core-collapse, in good agreement
with the ``five months'' estimate, given at the end of 
Section III (which was conservative in the sense that 
slightly more time was allotted for Bright Spot 2 to 
reach a projected distance of 160 mas). Once the delay,
between the initiation of high velocity jets (which
affects the luminosity of the common envelope star), and 
core-collapse, is known, aside from the size of the 
star, few other parameters, including the inclination 
wrt the Earth, should strongly affect it.

\begin{acknowledgments}

I am grateful for support for this work through the 
Los Alamos National Laboratory exploratory research 
and development grant no. 
20180352 ER, ''Scalable 
dielectric technology for VLF antennas.''  I also 
thank Andrea Schmidt and John Singleton for 
discussions and encouragement, and Hui Li and the 
rest of T-2 and T Division for support.

\end{acknowledgments}


\begin{thebibliography}{15}

\bibitem{PJ89} Podsiadlowski, Ph., \& Joss, P. C., 
``An alternative binary model for SN1987A,'' Nature 
\textbf{338}, 401--403 (1989).

\bibitem{P91} Podsiadlowski, Ph., Fabian, A. C., \& 
Stevens, I. R., ``Origin of the Napoleon's hat nebula 
around SN1987A and implications for the progenitor,''
Nature \textbf{453}, 43--46 (1991).

\bibitem{M00a}John Middleditch, Jerome A. Kristian, 
William E. Kunkel, Kym M. Hill, Robert D. Watson,
Richard Lucinio, James N. Imamura, Thomas Y. 
Steiman-Cameron, Andrew Shearer, Raymond Butler,
Michael Redfern, \& Anthony C. Danks, ``Rapid 
Photometry of supernova 1987A: a 2.14 ms pulsar?''
NewA \textbf{5} (5), 243--283 (2000).  


\bibitem{M00c}John Middleditch, Jerome A. Kristian, 
William E. Kunkel, Kym M. Hill, Robert D. Watson, 
Richard Lucinio, James N. Imamura, Thomas Y. 
Steiman-Cameron, Scott M. Ransom, Andrew Shearer, 
Raymond Butler, Michael Redfern, \& Anthony C. 
Danks, ``A 2.14 ms Candidate Optical Pulsar in SN 
1987A'' (2000). ArXiv:astro-ph0010044

\bibitem{M18} John Middleditch, ``Disruption of 
supernovae and would-be `Direct Collapsars','' 
arXiv:1910.03789 (2019).

\bibitem{C95} Christopher J. Burrows, John Krist, 
Jeff J. Hester, et al., ``Hubble Space Telescope
Observations of the SN 1987A Triple Ring
Nebula,'' ApJ \textbf{452} (1), 680--684 (1995).

\bibitem{MP07} Tom Morris \& Philippe Podsiadlowski, 
``The Triple-Ring Nebula Around SN 1987A:
Fingerprint of a Binary Merger,'' Sci 
\textbf{315} (5815), 1103--1106 (2007).

\bibitem{C87} A. Cassatella, C. Fransson, J. Vant 
Santvoort, C. Gry, A. Talavera, W. Wamsteker, N. 
Panagia, "Spectral evolution of SN 1987 A in the 
far-ultraviolet,'' A\&A \textbf{177} (1-2), 
L29--L32 (1987).

\bibitem{HD87} R. W. Hanuschik, J. Dachs, ``The 
H-alpha velocity structure during the first month 
of SN 1987a in the LMC,'' A\&A \textbf{182} (1), 
L29--L32 (1987).

\bibitem{Pap89} C. Papaliolios, L. Keochlin, P. 
Nisenson, C. Standley, S. Heathcote, ``Asymmetry 
of the envelope of supernova 1987A,'' Nature
\textbf{338}, 565--566 (1989).

\bibitem{W02} L. Wang, J. C. Wheeler, P. Hoflich, 
A. Khokhlov, D. Baade, D. Branch, P. Challis, A. V. 
Filippenko, C. Fransson, P. Garnavich, R. P. 
Kirshner, P. Lundqvist, R. McCray, N. Panagia, 
C. S. J. Pun, M. M. Phillips, G. Sonneborn, N. B. 
Suntzeff, ``The Axisymmetric Ejecta of Supernova 
1987A,'' ApJ \textbf{579} (2), 671--677 (2002).

\bibitem{Nis87} Peter Nisenson, Costas Papaliolios, 
Marguerite Karovska, \& Robert Noyes, ``Detection 
of a very bright source close to the LMC supernova 
SN 1987A,'' ApJ \textbf{320} (2), L15--L18 (1987).

\bibitem{MMM} W.~P.~S.~Meikle, S. J. Matcher, \& 
B. L. Morgan, ``Speckle interferometric observations 
of supernova 1987A and of a bright associated 
source,'' Nature \textbf{329}, 608--611 (1987).

\bibitem{NP99} Peter Nisenson, \& Cos Papaliolios, 
``A Second Bright Source Detected near SN 1987A,'' 
ApJ \textbf{518} (1), L29--L32, (1999).

\bibitem{HS90} Mario Hamuy, \& Nicholas B. Suntzeff, 
``SN 1987A in the LMC. III - UBVRI photometry at 
Cerro Tololo,'' AJ \textbf{99}, 1146--1158 (1990).

\bibitem{W87} W. Wamsteker, Nino Panagia, M. 
Barylak, et al., ``Early observations of supernova 
1987 A with the International Ultraviolet 
Explorer (IUE),'' A\&A \textbf{177} (1-2), 
L21--L24 (1987).

\bibitem{M12} John Middleditch, ``Pulsar-Driven 
Jets in Supernovae, Gamma-Ray Bursts, and the 
Universe,'' \textbf{Ad. Ast.} id. 898907, 26pp, \url{<http://www.hindawi.com/journals/aa/2012/898907>} 
(2012).


\bibitem{Men87} J. W. Menzies, R. M. Catchpole, 
G. Van Vuuren, et al., ``Spectroscopic and 
photometric observations of SN 1987a - The first 
50 days,'' MNRAS \textbf{277}, P39--P49 (1987).

\bibitem{Ar94} Houshang Ardavan, ``The Mechanism 
of Radiation in Pulsars,'' MNRAS \textbf{268} (2), 
361--392 (1994).

\bibitem{H98} Houshang Ardavan, ``Generation of 
focused, non-spherically decaying pulses of 
electromagnetic radiation,'' Phys. Rev. 
\textbf{58} (5), 6659--6684 (1998).

\bibitem{AR03} Houshang Ardavan, Arzhang 
Ardavan, \& John Singleton, ``Frequency 
spectrum of focused broadband pulses of 
electromagnetic radiation generated by 
polarization currents with superluminally 
rotating distribution patterns,'' JOSAA 
\textbf{20} (11), 2137--2155 (2003).

\bibitem{AR04} Houshang Ardavan, Arzhang 
Ardavan, \& John Singleton, ``Spectral and 
polarization characteristics of the 
nonspherically decaying radiation generated 
by polarization currents with superluminally 
rotating distribution patterns,'' JOSAA 
\textbf{21} (5), 858--872 (2004).

\bibitem{AR07} Houshang Ardavan, Arzhang 
Ardavan, \& John Singleton, ``Morphology of 
the nonspherically decaying radiation beam 
generated by a rotating superluminal source,'' 
JOSAA \textbf{24} (8), 2443--2458 (2007).

\bibitem{AR08} Houshang Ardavan, Arzhang 
Ardavan, John Singleton, Joseph Fasel, \& 
Andrea Schmidt, ``Spectral properties of the 
nonspherically decaying radiation generated 
by a rotating superluminal source,'' JOSAA 
\textbf{25} (3), 780--784 (2008).



\bibitem{BW20} Stefano Benetti, \& J. Craig 
Wheeler, AQ, in press (2018).


\bibitem{DH23} Dillon Z. Dong, \& Gregg Hallinan, ``A 
Flat-spectrum Radio Transient at 122 Mpc Consistent with 
an Emerging Pulsar Wind Nebula,'' ApJ \textbf{948}:119, 
25pp (2023).




\bibitem{jg22} W. V. Jacobson-Galan, L. Dessart, D. O. 
Jones, et al.,``Final Moments. I. Precursor Emission, 
Envelope Inflation, and Enhanced Mass Loss Preceding the 
Luminous Type II Supernova 2020tlf,'' ApJ \textbf{924} 
(2), 15 (2022).

\end{thebibliography}
\end{document}